\documentclass[twocolumn,aps,showpacs,prc]{revtex4-2}
\usepackage{epsfig}[dvips]
\begin{document}

\title{Radiative neutron capture reaction rates for stellar nucleosynthesis}

\author{ Vinay Singh$^{1}$, Debasis Bhowmick$^{2}$ and D. N. Basu$^{3}$}

\affiliation{Variable Energy Cyclotron Centre, 1/AF Bidhan Nagar, Kolkata 700064, INDIA}

\email[E-mail 1: ]{vsingh@vecc.gov.in}
\email[E-mail 2: ]{dbhowmick@vecc.gov.in}
\email[E-mail 3: ]{dnb@vecc.gov.in} 

\date{\today }

\begin{abstract}
	
    There is a high demand for nuclear data in multidisciplinary subject like nuclear astrophysics. The two areas of nuclear physics which are most clearly related to one another are stellar evolution and nucleosynthesis. The necessity for nuclear data for astrophysical applications puts experimental methods as well as reliability and predicative ability of current nuclear models to the test. Despite recent, considerable advances, there are still significant issues and mysteries. Only a few characteristics of nuclear astrophysics are covered in the current work which include $^{20}$Ne(n,$\gamma$)$^{21}$Ne, $^{52}$Fe(n,$\gamma$)$^{53}$Fe, $^{53}$Fe(n,$\gamma$)$^{54}$Fe, $^{54}$Fe(n,$\gamma$)$^{55}$Fe and $^{55}$Fe(n,$\gamma$)$^{56}$Fe reactions which are important in stellar nucleosynthesis. The reaction rates are calculated using nuclear statistical model. These rates are subsequently fitted to polynomials of temperature T$_9$ in order to facilitate calculations for stellar nucleosynthesis. 
\end{abstract}

\maketitle

\noindent
\section{Introduction}
\label{section1}

    The stars produce every chemical element in the cosmos, excluding cosmic hydrogen and helium, both during their quiet and explosive phases. The chemical constituents of sun show contamination from star generations that are now extinct from several epochs before the Solar System was formed. The slow neutron capture process (also known as the s-process) and the fast neutron capture process (also known as the r-process) are the two main nucleosynthesis processes that give rise to elements heavier than iron. The intermediate neutron capture process (also known as the ``i-process") is connected to a third, less frequent, nucleosynthesis channel, the existence of which is still unknown. The p-process also produces a few proton-rich isotopes.

    The metals (i.e., elements with A$\ge$12) have gradually become more abundant over the chemical history of universe at the expense of cosmic hydrogen. According to the s- (slow) and r- (rapid) processes, the distribution of heavy elements (A$>$56) in the solar system exhibits an exponential decline that is overlaid on a sequence of double peaks \cite{Bu57}. The Coulomb barrier does not prevent the neutron capture process, which are mostly involved in the synthesis of heavier elements around and beyond iron. Because neutron capture cross sections often rise as energy decreases, this type of nucleosynthesis may be easily accomplished at low energies. 

    The most frequent neutron fluxes in stars fall into one of two categories: s-process (n$_n \sim$ 10$^7$ cm$^{-3}$) or r-process (n$_n$ $>$ 10$^{20}$ cm$^{-3}$). In the vast space between these two densities, one may theoretically anticipate coming across astrophysical neutron fluxes. Even if those occurrences are not generally acknowledged, they do not appear to be frequent, despite rising evidence in the last ten years supporting the requirement for such an intermediate neutron capture process (i-process n$_n \sim$ 10$^{14}$ cm$^{-3}$); \cite{Cr16,De19,Ch21}). Low mass Asymptotic Giant Branch (AGB) stars and quiescent stages of large star evolution have been categorically recognized as two stellar sites for the s-process \cite{Cr09,Cr11,Cr15} and dormant periods of the development of big stars \cite{Pi10}. On the other hand, the r-process is thought to take place during the late phases of the evolution of massive stars (single or binary), while the precise location is still up for question (Magneto-rotational Super-novae \cite{Re21}, Neutron star mergers (NSMs) \cite{Ka17}, and Collapsars \cite{Si19}).

    In the present work, a few neutron capture reactions $^{20}$Ne(n,$\gamma$)$^{21}$Ne, $^{52}$Fe(n,$\gamma$)$^{53}$Fe, $^{53}$Fe(n,$\gamma$)$^{54}$Fe, $^{54}$Fe(n,$\gamma$)$^{55}$Fe and $^{55}$Fe(n,$\gamma$)$^{56}$Fe which are important in stellar nucleosynthesis have been studied. The reaction cross sections have been calculated using Hauser-Feshbach statistical model. These cross sections have been convoluted with the Maxwell-Boltzmann energy distribution to get the thermonuclear reaction rates. Subsequently, in order to facilitate stellar nucleosynthesis calculations, these reaction rates have been fitted to polynomials of temperature T$_9$ (10$^9$ K).  
		
\noindent
\section{Hauser-Feshbach statistical model}
\label{section2}

    The fusion cross sections can be convoluted with the Maxwell-Boltzmann energy distribution to get the thermonuclear reaction rates. These cross sections over the requisite energy range, do vary by quite a few orders of magnitude. Laboratory experiments can be used to determine the low energy fusion cross sections $\sigma$, some of which are not sufficiently well characterized. However, there are other situations, particularly those involving the weak interaction, where there are no experimental data available and one must solely rely on theoretical calculations \cite{Ad11}, like the solar p-p chain deuterium formation via the fundamental p$+$p fusion reaction. The various approximations utilized affect the theoretical estimation for any thermonuclear reaction rate calculation. The measured values of the cross sections are influenced by a number of factors. The network computations for primordial or stellar nucleosynthesis must take into consideration the Maxwellian-averaged thermonuclear reaction rates.

    The reaction rates calculated in the present work are not independent of temperature. In addition to conventional nuclear physics calculations, the computer programme TALYS \cite{Talys} enables thorough estimates of astrophysical reaction rates. The assumption of thermodynamic equilibrium is approximately valid in the interior of stars, where nuclei can exist in both the ground and excited states. Maxwellian-averaged reaction rates are facilitated by this presumption combined with cross sections computed using the compound nucleus model for various excited states. This is a crucial input for models of star evolution. Astrophysical calculations primarily require the nuclear reaction rates, which are often assessed using the statistical model \cite{Rauscher97,Rauscher10}. Calculations of the stellar reaction rate have frequently been performed in the past \cite{Rauscher00,Rauscher01}. However, TALYS has added some new and significant aspects to these Hauser-Feshbach statistical model (HF) estimates \cite{Ha52}. Along with the coherent inclusion of the fission channel, other features include multi-particle emission, competition between all open channels, in-depth width fluctuation corrections, coupled channel description in the case of deformed nuclei, and parity-dependent level densities. The nuclear models are also adjusted for the experimental data using several methods, such as the level densities, the E1 resonance strength, or the photoabsorption data.

\subsection{Neutron Capture reaction cross section}

    The projectile and target nuclei fuse together to form a compound nucleus in the low energy region. The spin $J$ and parity $\Pi$ can have a variety of values, however the total energy $E^{tot}$ is fixed due to energy conservation. The following conservation laws are followed by the reaction:
$$
E_{a}+S_{a}\ =\ E_{a'}+E_{x}+S_{a'}=E^{tot},~~~~{\rm energy~conservation},  
$$
$$
s+I+l\ =\ s'+I'+l'=J,~~~~{\rm angular~momentum~conservation},
$$
\begin{center}
  $\pi_{0}\Pi_{0}(-1)^{l}\ = \pi_f\Pi_f(-1)^{l'}=\Pi,~~~~{\rm parity~conservation}.$
\end{center}
The formula for binary cross section, assuming the compound nucleus model, is given by

\begin{eqnarray}
\sigma_{\alpha\alpha'}^{comp}\ &&=\ D^{comp}\frac{\pi}{k^{2}}\sum_{J=mod(I+s,1)}^{l_{\max}+I+s}
\sum_{\Pi=-1}^{1}\frac{2J+1}{(2I+1)(2s+1)}  \nonumber\\
&&\sum_{j=|J-I|}^{J+I}\sum_{l=|j-s|}^{j+s}\sum_{j'=|J-I'|}^{J+I'}\sum_{l'=|j'-s'|}^{j'+s'} \delta_{\pi}(\alpha)\delta_{\pi}(\alpha')
\end{eqnarray}

\begin{center}
   $\displaystyle \times\ \frac{T_{\alpha lj}^{J}(E_{a})\langle T_{\alpha' l'j'}^{J}(E_{a'})\rangle}{\sum_{\alpha'',l'',j''}\delta_{\pi}(\alpha'')\langle T_{\alpha''l''j''}^{J}(E_{a''})\rangle}W_{\alpha lj\alpha'l'j'}^{J}$
\end{center}
\noindent
where

$E_{a}=$ the energy of the projectile 

$l=$ the orbital angular momentum of the projectile

$s=$ the spin of the projectile

$j=$ the total angular momentum of the projectile

$\pi_{0}=$ the parity of the projectile

$\delta_{\pi}(\alpha)=\left\{ \begin{array}{ll}
1 ~~&{\rm if}~~(-1)^{l}\pi_{0}\Pi_{0}=\Pi \\ 
 0~~& {\rm otherwise}
\end{array}\right.$ 

$\alpha=$ the designation of the channel for the initial projectile-target system:

$\alpha=\{a,\ s,\ E_{a},\ E_{x}^{0},\ I,\ \Pi_{0}\}$, where $a$ and $E_{x}^{0}$ are the type of the projectile and  the excitation energy (which is zero usually) of the target nucleus, respectively 

$l_{\max}=$ the maximum l-value of the projectile

$S_{a}=$ the separation energy

$E_{a'} =$  the energy of the ejectile

$l'=$ the orbital angular momentum of the ejectile

$s'=$ the spin of the ejectile

$j'=$ the total angular momentum of the ejectile 

$\pi_{f}=$ the parity of the ejectile

$\delta_{\pi}(\alpha')=\left\{ \begin{array}{ll}
1 ~~&{\rm if}~~(-1)^{l'}\pi_{f}\Pi_{f}=\Pi \\ 
 0~~& {\rm otherwise}
\end{array}\right.$ 

$\alpha'=$ the designation of channel for the ejectile-residual nucleus final system:

$\alpha'=\{a',\ s',\ E_{a'},\ E_{x},\ I',\ \Pi_f\}$, where $a'$ and  $E_{x}$ are the type of the ejectile and the residual nucleus excitation energy, respectively

$I=$ the spin of target nucleus

$\Pi_{0}=$ the parity of target nucleus

$I'=$ the spin of residual nucleus

$\Pi_f=$ the parity of residual nucleus

$J=$ the total angular momentum of the compound system

$\Pi=$ the parity of the compound system

$D^{comp}=$ the depletion factor so as to take into account for pre-equilibrium and direct effects

$k=$ the wave number of the relative motion

$T=$ the transmission coefficient

$W=$ the correction factor for width fluctuation (WFC).

\begin{figure}[ht!]
\vspace{0.0cm}
\eject\centerline{\epsfig{file=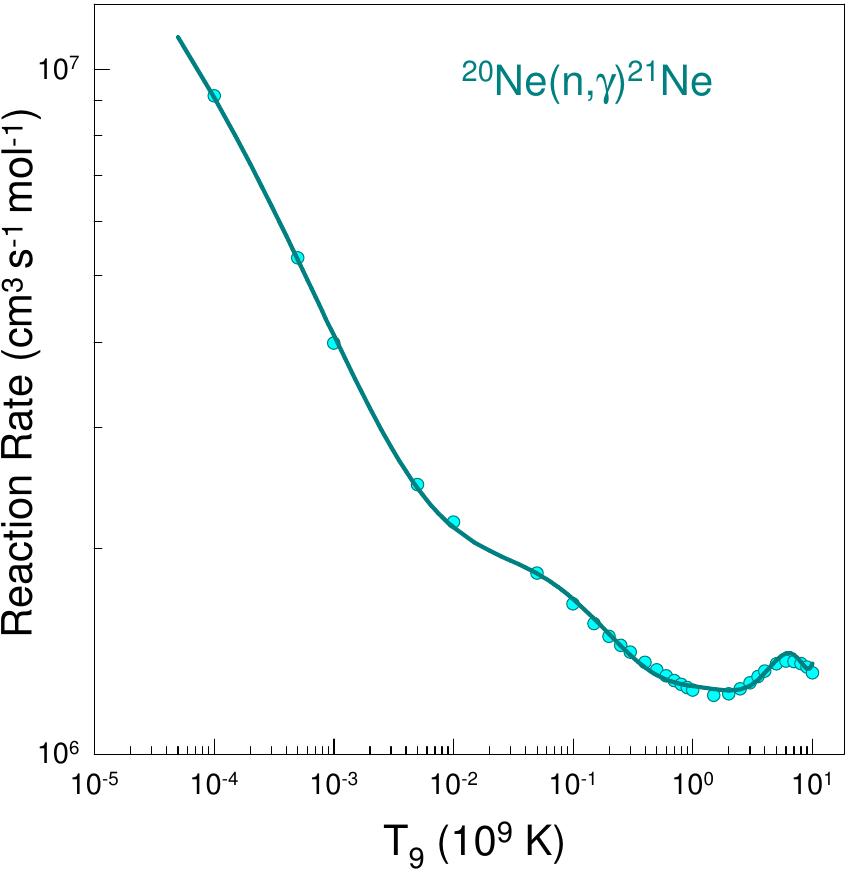,height=7.7cm,width=7.7cm}}
\caption
{The Hauser-Feshbach estimates of $^{20}$Ne(n,$\gamma$)$^{21}$Ne thermonuclear reaction rate versus temperature T$_9$.}
\label{fig1}
\vspace{0.0cm}
\end{figure}
\noindent 

\subsection{ The rate equations for (n,$\gamma$) reactions }

    The target and projectile velocities follow Maxwell-Boltzmann distributions that correspond to the site having ionic plasma temperature $T$. For energies $E$ the Maxwell-Boltzmann energy distribution at a specified temperature $T$ and the cross section obtained by Eq.(1) can be folded to determine the astrophysical thermonuclear reaction rates. In addition, the target nuclei can be found in both the ground and excited states. The Maxwell-Boltzmann distribution describes the relative populations of different energy states of nuclei with excitation energies $E_{x}^{\mu}$ and spins $I^{\mu}$ in thermodynamic equilibrium. The superscript $\mu$ and incident $\alpha$ channel are utilized in the calculations that follow to differentiate between various excited states. The final expression for the effective $\alpha\rightarrow\alpha'$ entrance channel nuclear reaction rate takes into consideration the contributions from various target nuclei excited states which can be given by  

\begin{equation}
 N_{A}\langle\sigma v\rangle_{\alpha\alpha'}^{*}(T)=\left(\frac{8}{\pi m}\right)^{1/2}\frac{N_{A}}{(kT)^{3/2}G(T)}\times\ 
\end{equation}

\begin{center}
   $\displaystyle  \int_{0}^{\infty}\sum_{\mu}\frac{(2I^{\mu}+1)}{(2I^{0}+1)}\sigma_{\alpha\alpha'}^{\mu} (E)E\exp\left(-\frac{E+E_{x}^{\mu}}{kT}\right)dE,$
\end{center}
where $N_{A}$, $m$ and $k$ and are, respectively, the Avogadro number, the reduced mass in the $\alpha$ channel and the Boltzmann constant and 

\begin{center}
  $G(T)=\displaystyle \sum_{\mu}(2I^{\mu}+1)/(2I^{0}+1)\exp(-E_{x}^{\mu}/kT)$
\end{center}
is the normalized partition function that depends upon temperature. The reverse reaction cross sections or rates can also be calculated using the reciprocity theorem \cite{Ho76}. 

\begin{figure}[ht!]
\vspace{0.0cm}
\eject\centerline{\epsfig{file=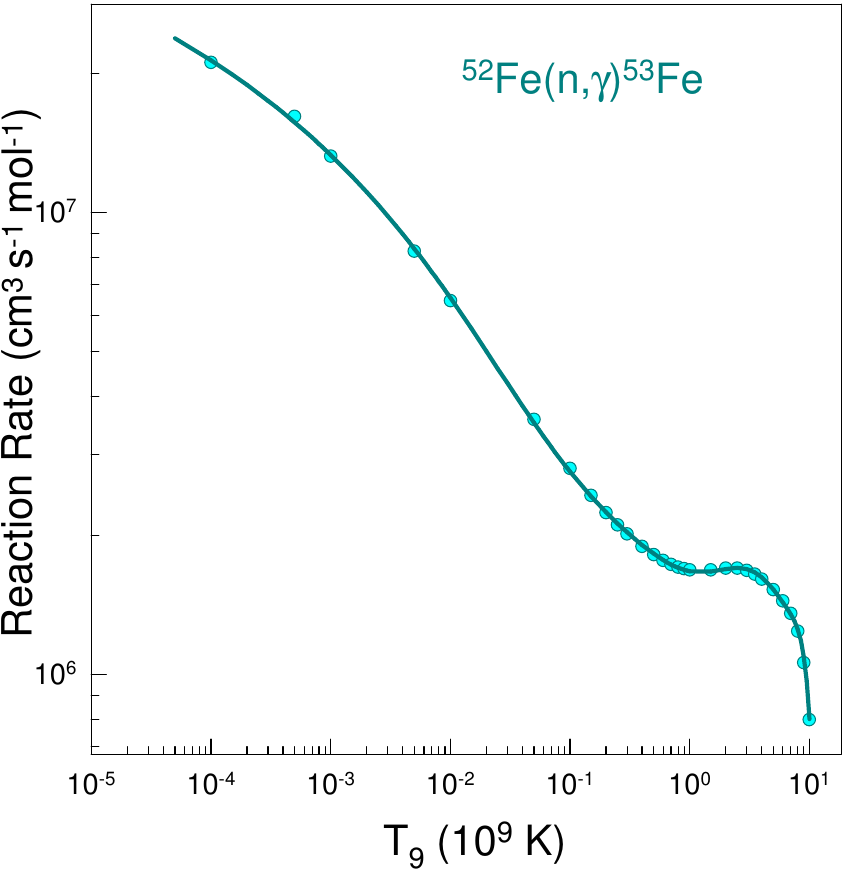,height=7.7cm,width=7.7cm}}
\caption
{The Hauser-Feshbach estimates of $^{52}$Fe(n,$\gamma$)$^{53}$Fe thermonuclear reaction rate versus temperature T$_9$.}
\label{fig2}
\vspace{0.0cm}
\end{figure}
\noindent
    
\noindent
\section{Calculations and Results}
\label{section3}

    Using the Hauser-Feshbach statistical model computations the radiative neutron capture (n,$\gamma$) reaction cross sections have been determined theoretically for $^{20}$Ne(n,$\gamma$)$^{21}$Ne, $^{52}$Fe(n,$\gamma$)$^{53}$Fe, $^{53}$Fe(n,$\gamma$)$^{54}$Fe, $^{54}$Fe(n,$\gamma$)$^{55}$Fe and $^{55}$Fe(n,$\gamma$)$^{56}$Fe reactions which are of astrophysical significance and important in stellar nucleosynthesis. The latest level density based on temperature-dependent Hartree-Fock-Bogolyubov calculations utilizing the Gogny force and the Brink-Axel Lorentzian for the gamma-ray strength function \cite{Talys} have been chosen for performing the present computations. These excitation functions highlight the variations of (n,$\gamma$) cross sections with energy and show a different energy dependence than ${\rm 1/E_n^{1/2}}$ behavior valid at very low energies in the thermal domain. The energy variation of radiative capture cross section is more expeditious in the astrophysical realm of energy range of 10 keV to 1.2 MeV than in the thermal domain. 

\begin{figure}[ht!]
\vspace{0.0cm}
\eject\centerline{\epsfig{file=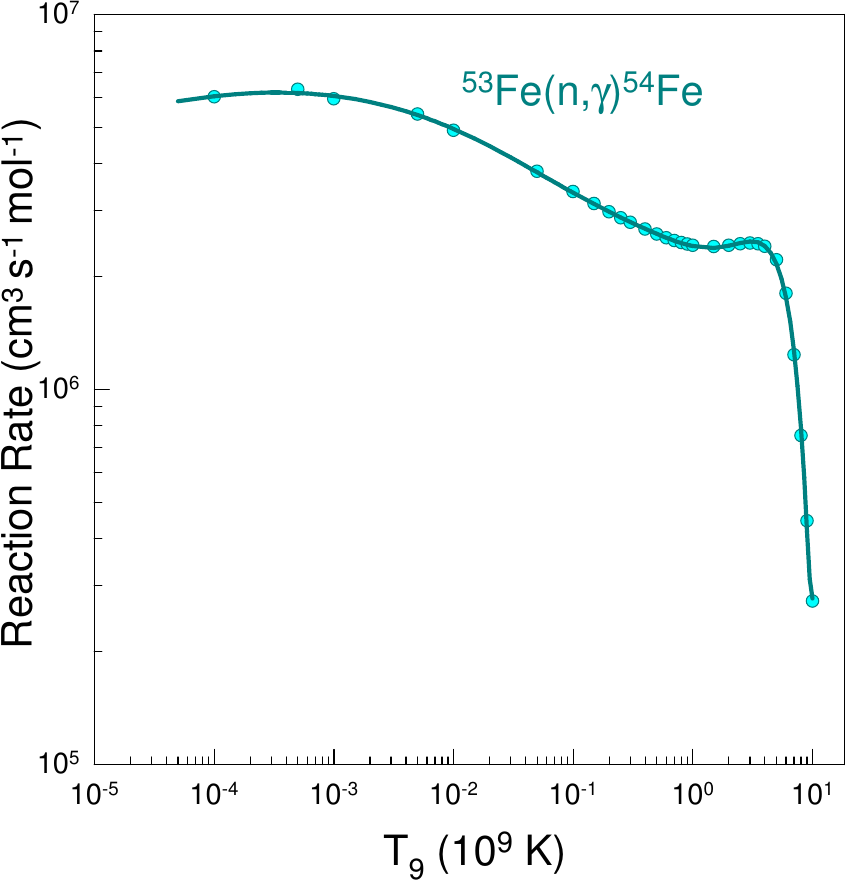,height=7.7cm,width=7.7cm}}
\caption
{The Hauser-Feshbach estimates of $^{53}$Fe(n,$\gamma$)$^{54}$Fe thermonuclear reaction rate versus temperature T$_9$.}
\label{fig3}
\vspace{0.0cm}
\end{figure}
\noindent

    In a stellar plasma, the kinetic energy available to nuclei is that of their thermal motion. Hence, reactions initiated by this motion are called thermonuclear reactions. The nuclei in a stellar plasma move non-relativistically and are non-degenerate at temperatures in the range of 10$^9$ K (T$_9$) ubiquitous in these environments. In stellar interiors, nuclides not only exist in their ground states but also in different thermally excited states and a thermodynamic equilibrium holds locally to a very good approximation. Therefore, velocities of nuclei can be described by a Maxwell-Boltzmann distribution. For the explanation of many processes undergoing in these extreme conditions, the nuclear reaction cross sections and their convolution with the Maxwell-Boltzmann distribution of energies are crucial \cite{Bu57,Fo64,Cl83}. The extremely high density and temperature are prevalent in main-sequence and compact stars that are in their late stages of evolution. Nuclear explosions are produced by exothermic fusion reactions in the cores of massive accreting white dwarfs (type Ia supernovae \cite{Ni97,Ho06}), the surface layers of accreting neutron stars (type I X-ray bursts and superbursts \cite{St06,Sc03,Cu06,Gu07}) and the surface layers of accreting white dwarfs (nova events). For the purpose of characterizing these astrophysical phenomena, it becomes imperative to have a precise understanding of the rates of thermonuclear reactions as determined by convoluting the Maxwell-Boltzmann distribution of energies with energy dependent cross sections. The integral \cite{Ad11,Fo67,Bo08} outlined below can be used to represent the Maxwellian-averaged thermonuclear reaction rate per particle pair $<\sigma v>$ at temperature T:  

\begin{equation}
 <\sigma v> = \Big[\frac{8}{\pi m (k T)^3 } \Big]^{1/2} \int \sigma(E) E \exp(-E/k T) dE,
\label{seqn3}
\end{equation}
\noindent
where $v$, $m$, $E$ and $k$ are, repectively, the relative velocity, the reduced mass of the reacting nuclei, the energy in the centre-of-mass system and Boltzmann constant. Hence, the thermonuclear reaction rate between two interacting nuclei is: $r_{12}=\frac{n_1n_2}{1+\delta_{12}}<\sigma v>$ with $n_i$ representing the number density of nuclei of types `$i$' where for preventing double counting in case of identical particles Kronecker delta $\delta_{12}$ has been used. The quantity $N_A<\sigma v>$ represents the reaction rate with $N_A$ being the Avogadro number which is equal to 6.023$\times 10^{23}$. In Figs.1-5, the plots of the (n,$\gamma$) reaction rates as functions of T$_9$ for the reactions $^{20}$Ne(n,$\gamma$)$^{21}$Ne, $^{52}$Fe(n,$\gamma$)$^{53}$Fe, $^{53}$Fe(n,$\gamma$)$^{54}$Fe, $^{54}$Fe(n,$\gamma$)$^{55}$Fe and $^{55}$Fe(n,$\gamma$)$^{56}$Fe expressed in units of cm$^3$mol$^{-1}$s$^{-1}$ have been shown by solid dots.

    The microscopic nuclear inputs used for the radiative neutron capture reaction cross section and reaction rate calculations have been specified in a suitable tabular format in Table-I  below.      
\begin{figure}[ht!]
\vspace{0.0cm}
\eject\centerline{\epsfig{file=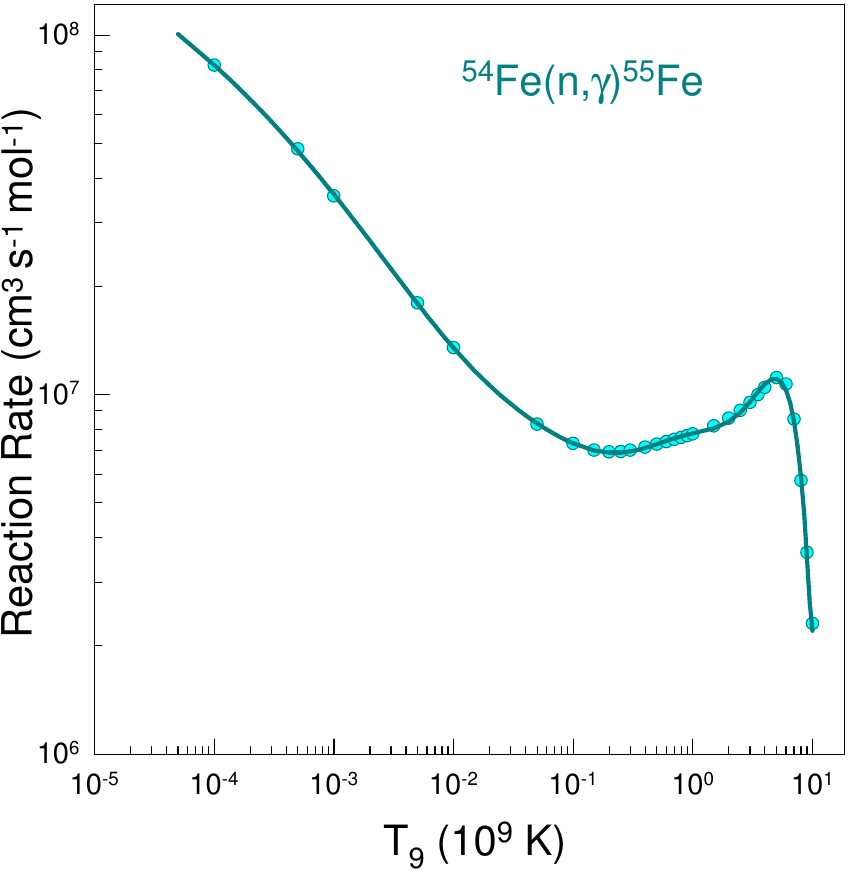,height=7.7cm,width=7.7cm}}
\caption
{The Hauser-Feshbach estimates of $^{54}$Fe(n,$\gamma$)$^{55}$Fe thermonuclear reaction rate versus temperature T$_9$.}
\label{fig4}
\vspace{0.0cm}
\end{figure}
\noindent   
    
\begin{table}[htbp]
\centering
\caption{Primary inputs used in Hauser-Feshbach statistical model reaction cross section and rate calculations.}
\vspace{0.2cm}
\begin{tabular}{||c c||}
\hline 
\hline
Physical quantity&  TALYS   \\ 
Model used&    Option    \\
\hline
\hline
Reaction &(n,$\gamma$)  \\ 

\hline
Nuclear interaction &Optical model potential  \\ 

\hline
Cross section calculation &\\
Energy range &10 keV to 1.2 MeV  \\ 

\hline
Reaction rate calculation & \\
Temperature range &10$^{-4}$ T$_9$ to 10 T$_9$  \\ 
Equivalent energy range &8.6173 eV to 0.86173 MeV  \\ 

\hline
Level density &ldmodel 6   \\ 
Temperature dependent &  \\ 
Hartree-Fock-Bogolyubov&  \\ 
using Gogny force&  \\ 

\hline
All level density parameters   &asys n     \\
from systematics&      \\

\hline
E1 $\gamma$-ray strength function &strength 2    \\
Brink-Axel Lorentzian&     \\ 

\hline
Normalization factor for   &gnorm -1     \\
$\gamma$-ray transmission coefficient&      \\

\hline
\hline
\end{tabular}
\label{table1} 
\end{table}
\noindent 		   

\begin{table*}[ht!]
\vspace{0.0cm}
\centering
\caption{\label{tab:table1} The co-efficients of fitting of thermonuclear reaction rate to the polynomial in temperature T$_9$.}
\begin{tabular}{cccccccccc}
\hline
\hline

Reaction&a$_0$&a$_1$&a$_2$&a$_3$&a$_4$&a$_5$&a$_6$&a$_7$&a$_8$ \\ \hline
$^{20}$Ne(n,$\gamma$)$^{21}$Ne& -0.386600&0.136663&-0.273377&0.364081&-0.298545&0.145169&-0.382799&0.420515&-0.104934  \\
&$\times 10^8$&$\times 10^9$&$\times 10^9$&$\times 10^9$&$\times 10^9$&$\times 10^9$&$\times 10^8$&$\times 10^7$&$\times 10^8$ \\
&&&&&&&&& \\

$^{52}$Fe(n,$\gamma$)$^{53}$Fe &-0.113774&-0.898350&0.109944&-0.212525&0.2117889&-0.117009&0.338290&-0.399156&-0.823436 \\
&$\times 10^8$&$\times 10^7$&$\times 10^9$&$\times 10^9$&$\times 10^9$&$\times 10^9$&$\times 10^8$&$\times 10^7$&$\times 10^7$ \\
&&&&&&&&& \\

$^{53}$Fe(n,$\gamma$)$^{54}$Fe&
0.153882&-0.443333&0.751030&-0.722473&0.315010&0.129019&-0.552535&0.123653&0.185681 \\
&$\times 10^8$&$\times 10^8$&$\times 10^8$&$\times 10^8$&$\times 10^8$&$\times 10^7$&$\times 10^7$&$\times 10^7$&$\times 10^7$ \\
&&&&&&&&& \\

$^{54}$Fe(n,$\gamma$)$^{55}$Fe& 
-0.309640&0.101116&-0.200123&0.284575&-0.256504&0.137874&-0.399704&0.477580&-0.873182 \\
&$\times 10^9$&$\times 10^{10}$&$\times 10^{10}$&$\times 10^{10}$&$\times 10^{10}$&$\times 10^{10}$&$\times 10^9$&$\times 10^8$&$\times 10^8$ \\
&&&&&&&&& \\

$^{55}$Fe(n,$\gamma$)$^{56}$Fe&
-0.843064&0.125863&0.798242&-0.357322&0.423271&-0.244802&0.697564&-0.779767&-0.359015 \\
&$\times 10^9$&$\times 10^{10}$&$\times 10^9$&$\times 10^{10}$&$\times 10^{10}$&$\times 10^{10}$&$\times 10^9$&$\times 10^8$&$\times 10^8$ \\

\hline
\hline 
\end{tabular} 
\vspace{0.0cm}
\end{table*}
\noindent        

\begin{figure}[htbp]
\vspace{2.95cm}
\eject\centerline{\epsfig{file=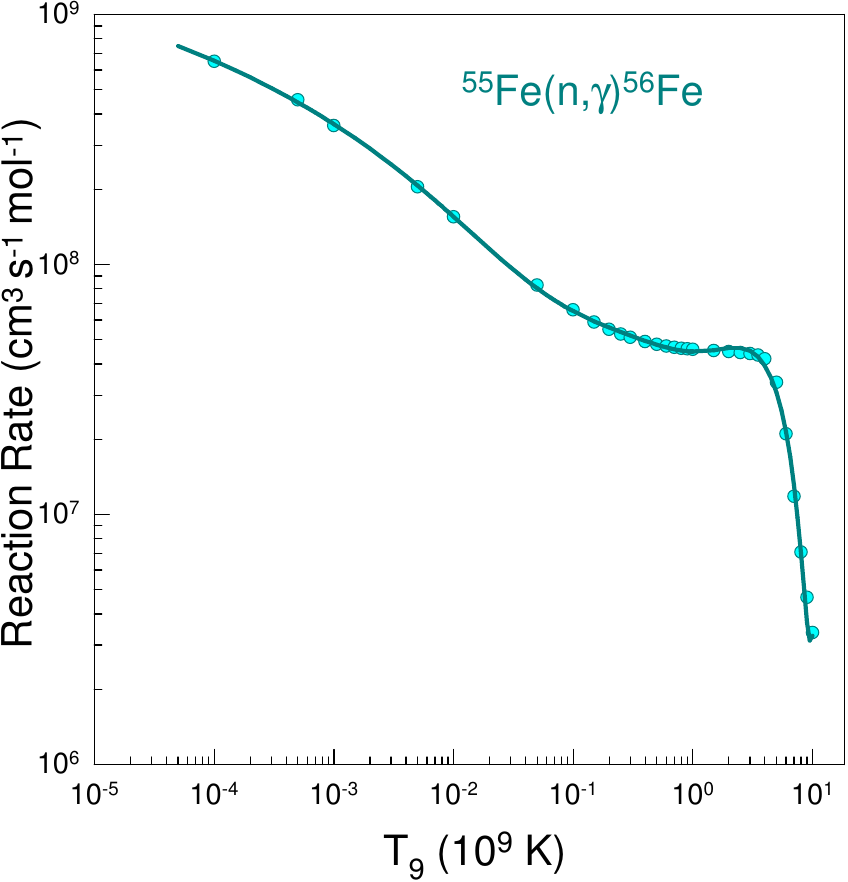,height=7.7cm,width=7.7cm}}
\caption
{The Hauser-Feshbach estimates of $^{55}$Fe(n,$\gamma$)$^{56}$Fe thermonuclear reaction rate versus temperature T$_9$.}
\label{fig5}
\vspace{0.0cm}
\end{figure}
\noindent 

    The flag `asys n' opted here is for using all level density parameters from systematics and for neglecting the connection between the level density parameter `a' and the s-wave resonance spacing `D$_0$', even if an experimental value is available for the latter. The normalisation factor for $\gamma$-ray transmission coefficient `gnorm' is an adjustable parameter that can be used to scale the (n, $\gamma$) cross sections. To enforce automatic normalization `gnorm'  has been chosen as -1. As the flag `localomp' has not been used, whenever enough experimental scattering data of a certain nucleus are available, a local optical model potential (OMP) can be constructed. The local OMP parameters are automatically obtained by TALYS from the database containing nuclear structure and model parameters. In case the database does not have a local OMP representation, the existing global OMPs are utilized. Furthermore, as there is no usage of the `massmodel' or `expmass' flags, theoretical mass models are employed only if there is no access to an experimental mass. 
    
    Investigations have been made for a few crucial reactions \cite{Su14} that could have a significant impact on the elemental abundances in stellar nucleosynthesis. In past, studies \cite{Fo88,Ma89,Sm93,An99,An04} have been conducted on reaction rates. All other reaction rates have temperature dependence, with the exception of a few neutron-induced reactions. In the thermal domain, the neutron capture cross sections exhibit $\approx 1/v$ behaviour at very low energies. As a result, one can see right away from Eq.(3), using $\sigma(E) \propto E^{-1/2}$, that at thermal energy, reaction rates are essentially constant with regard to temperature. However, this finding only holds true for very low energy neutrons ($\sim$ 0.025 eV) when energies are $\sim$ eV and lower. On the other hand, the capture cross sections for neutron-induced processes can be best characterized as $\sigma(E)=\frac{R(E)}{v}$ \cite{Bl55}, where $R(E)$ softly changes as a function of energy \cite{Mu10}, at energies in the range of astrophysical interest. As a result, it is comparable to the astrophysical S-factor, and one would anticipate that $\langle\sigma v\rangle$ would vary with temperature. The thermonuclear reaction rates  $N_A<\sigma v>$ have been fitted to the polynomials of temperature T$_9$ of the form  
    
\begin{eqnarray}
N_A<\sigma v> = a_0 + a_1 T_9^{1/3} + a_2 T_9^{2/3} + a_3 T_9\nonumber\\
  + a_4 T_9^{4/3} + a_5 T_9^{5/3} + a_6 T_9^2  \nonumber\\
  + a_7 T_9^{7/3} + a_8 {\rm log_{10}}T_9  
\label{seqn4}
\end{eqnarray}
\noindent    
and have been presented in Figs.1-5 by continuous lines. The co-efficients of fitting of thermonuclear reaction rate to the polynomial in temperature T$_9$ have been listed in Table-II.  
	    
\noindent
\section{Summary and conclusion}
\label{section4}

    This work estimates the thermonuclear (n,$\gamma$) reaction rates of astrophysical importance for $^{20}$Ne(n,$\gamma$)$^{21}$Ne, $^{52}$Fe(n,$\gamma$)$^{53}$Fe, $^{53}$Fe(n,$\gamma$)$^{54}$Fe, $^{54}$Fe(n,$\gamma$)$^{55}$Fe and $^{55}$Fe(n,$\gamma$)$^{56}$Fe reactions using the Hauser-Feshbach statistical model. A few important thermonuclear reactions \cite{Su14} which may have significant impact on the chemical abundances in the region near $^{56}$Fe have been explored. The $^{56}$Fe nucleus formed at the endpoint of thermonuclear burning is the most energetically favorable one at the low densities. In order to facilitate the stellar nucleosynthesis computations, these rates are then fitted to polynomials of temperature T$_9$.  

    To conclude, the present investigation have important bearings on the relative abundance of the elements involved in the stellar nucleosynthesis. In order to study the consequence of the present results on the elemental abundances and isotopic ratios, a comprehensive and detailed full reaction network calculations at evolving neutron densities needs to be performed in future. The lack of sufficient experimental data is one of the biggest impediments in constraining the (n,$\gamma$) reaction rates near the mass region fifty-six which can probably be acquired using new experimental techniques such as the surrogate method \cite{Koz12} or the beta-Oslo method \cite{Sp14}. This would also facilitate exclusion or establishing certain model inputs in future theoretical calculations.
\vspace{-0.7cm}
\begin{acknowledgements}
\vspace{-0.4cm}
    One of the authors (DNB) acknowledges support from Science and Engineering Research Board, Department of Science and Technology, Government of India, through Grant No. CRG/2021/007333.
\vspace{-0.2cm}
\end{acknowledgements}


\vspace{-0.4cm}

\noindent
\begin{thebibliography}{99}

\bibitem{Bu57} E. M. Burbidge, G. R. Burbidge, W. A. Fowler and F. Hoyle, Reviews of Modern Physics {\bf 29}, 547 (1957).

\bibitem{Cr16} S. Cristallo, D. Karinkuzhi, A. Goswami, L. Piersanti and D. Gobrecht, ApJ {\bf 833}, 181 (2016), arXiv:1610.05475.

\bibitem{De19} P. A. Denissenkov, F. Herwig, P. Woodward, R. Andrassy, M. Pignatari and S. Jones, MNRAS {\bf 488}, 4258 (2019), arXiv:1809.03666.

\bibitem{Ch21} A. Choplin, L. Siess and S. Goriely, A$\&$A {\bf 648}, A119 (2021), arXiv:2102.08840.

\bibitem{Cr09} S. Cristallo, O. Straniero, R. Gallino, L. Piersanti, I. Domínguez and M. T. Lederer, ApJ {\bf 696}, 797 (2009), arXiv:0902.0243.

\bibitem{Cr11} S. Cristallo, L. Piersanti, O. Straniero, R. Gallino, I. Domínguez, C. Abia, G. Di Rico, M. Quintini and S. Bisterzo, ApJS {\bf 197}, 17 (2011), arXiv:1109.1176.

\bibitem{Cr15} S. Cristallo, O. Straniero, L. Piersanti and D. Gobrecht, ApJS {\bf 219}, 40 (2015), arXiv:1507.07338.

\bibitem{Pi10} M. Pignatari, R. Gallino, M. Heil, M. Wiescher, F. Käppeler, F. Herwig and S. Bisterzo, ApJ {\bf 710}, 1557 (2010).

\bibitem{Re21} M. Reichert, M. Obergaulinger, M. Eichler, M. Á. Aloy and A. Arcones, MNRAS {\bf 501}, 5733(2021), arXiv:2010.02227.

\bibitem{Ka17} D. Kasen, B. Metzger, J. Barnes, E. Quataert and E. Ramirez-Ruiz, Nature {\bf 551}, 80 (2017), arXiv:1710.05463.

\bibitem{Si19} D. M. Siegel, J. Barnes and B. D. Metzger, Nature {\bf 569}, 241 (2019), arXiv:1810.00098.

\bibitem{Ad11} E. G. Adelberger et al., Rev. Mod. Phys. {\bf 83}, 195 (2011).

\bibitem{Talys} A. J. Koning, S. Hilaire, and M. C. Duijvestijn. In Proceedings of the International Conference on Nuclear Data for Science and Technology, 22-27 April 2007, Nice, France. 2008. pp. 211-214. Arjan Koning, Stephane Hilaire and Stephane Goriely, {\bf TALYS-1.8} A nuclear reaction program, December 26, (2015).

\bibitem{Rauscher97} T. Rauscher, F.-K. Thielemann and K.-L. Kratz, Phys. Rev. {\bf C 56}, 1613 (1997).

\bibitem{Rauscher10} T. Rauscher, Phys. Rev. {\bf C 81}, 045807 (2010).

\bibitem{Rauscher00} T. Rauscher and F.-K. Thielemann, Atom. Data Nucl. Data Tables {\bf 75}, 1 (2000).

\bibitem{Rauscher01} T. Rauscher and F.-K. Thielemann, Atom. Data Nucl. Data Tables {\bf 79}, 47 (2001). 

\bibitem{Ha52} W. Hauser and H. Feshbach, Phys. Rev. {\bf 87}, 366 (1952).

\bibitem{Ho76} J. A. Holmes, S. E. Woosley, W. A. Fowler and B. A. Zimmerman, Atom. Data Nucl. Data Tables {\bf 18}, 306 (1976).

\bibitem{Fo64} W. A. Fowler and F. Hoyle, Astrophys. J. Suppl. {\bf 9}, 201 (1964); Appendix C.

\bibitem{Cl83} D. D. Clayton, {\it Principles of Stellar Evolution and Nucleosynthesis} (University of Chicago Press, Chicago, 1983).

\bibitem{Ni97} J. C. Niemeyer, S. E. Woosley, Astrophys. J. {\bf 475}, 740 (1997).

\bibitem{Ho06} P. H\"oflich, Nucl. Phys. {\bf A 777}, 579 (2006).

\bibitem{St06} T. Strohmayer, L. Bildsten, {\it New views of thermonuclear bursts} in: W. H. G. Lewin, M. Van der Klis (Eds.), {\it Compact Stellar X-ray Sources}, (Cambridge University Press, Cambridge, London.), 113 (2006).

\bibitem{Sc03}  H. Schatz, L. Bildsten, A. Cumming, Astrophys. J. {\bf 583}, L87 (2003).

\bibitem{Cu06}  A. Cumming, J. Macbeth, J.J.M. in t Zand, D. Page, Astrophys. J. {\bf 646}, 429 (2006).

\bibitem{Gu07}  S. Gupta, E. F. Brown, H. Schatz, P. M\"oller, K.-L. Kratz, Astrophys. J. {\bf 662}, 1188 (2007).

\bibitem{Fo67} W. A. Fowler, G. R. Caughlan and B. A. Zimmerman, Ann. Rev. Astron. Astrophys. {\bf 5} (1967) 525.

\bibitem{Bo08} R. N. Boyd, {\it An Introduction to Nuclear Astrophysics} (University of Chicago, Chicago, 2008), 1st ed.

\bibitem{Su14} R. Surman, M. Mumpower, R. Sinclair, K. L. Jones, W. R. Hix and G. C. McLaughlin, AIP Advances {\bf 4}, 041008 (2014).

\bibitem{Fo88} G. R. Caughlan and W. A. Fowler, Atom. Data Nucl. Data Tables {\bf 40}, 283 (1988).

\bibitem{Ma89} R. A. Malaney and W. A. Fowler, Astrophys. J. {\bf 345}, L5 (1989).

\bibitem{Sm93} M. S. Smith, L. H. Kawano and R. A. Malaney, Astrophys. J. Suppl. {\bf 85}, 219 (1993).

\bibitem{An99} C. Angulo et al., Nucl. Phys. {\bf A 656}, 3 (1999).

\bibitem{An04} P. Descouvemont, A. Adahchour, C. Angulo, A. Coc and E. Vangioni-Flam, Atom. Data Nucl. Data Tables {\bf 88}, 203 (2004).

\bibitem{Bl55} J. M. Blatt and V. F. Weisskopf, {\it Theoretical Nuclear Physics} (John Wiley $\&$ Sons, New York; Chapman $\&$ Hall Limited, London.) 

\bibitem{Mu10} Tapan Mukhopadhyay, Joydev Lahiri and D. N. Basu, Phys. Rev. {\bf C 82}, 044613 (2010); {\it ibid} Phys. Rev. {\bf C 83}, 067603 (2011).

\bibitem{Koz12} Kozub et al., Phys. Rev. Lett. {\bf 109}, 172501 (2012).

\bibitem{Sp14} A. Spyrou et al., Phys. Rev. Lett. {\bf 113}, 232502 (2014).

\end{thebibliography}
\end{document}